\documentclass[12pt]{article}
\usepackage{graphicx}
\begin{document}
\bigskip
\centerline{\large \bf
Concentration Gradient, Diffusion, and Flow Through}
\centerline{\large \bf 
Open Porous Medium Near Percolation Threshold} 
\centerline{\large \bf
via Computer Simulations} 
\bigskip
\bigskip
\bigskip
\centerline {\it R.B. Pandey$^{1,2}$, J.F. Gettrust$^1$, and 
D. Stauffer$^{3,4}$}
\bigskip
\bigskip

\centerline{$^1$Naval Research Laboratory}
\centerline{Stennis Space Center, MS 39529}
\bigskip
\centerline{$^2$Department of Physics and Astronomy}
\centerline{University of Southern Mississippi,
Hattiesburg, MS 39406-5046}
\bigskip
\centerline{$^3$Instituto de Fisica}
\centerline{Universidade Federal Fluminense, Av. Litoranea s/n, Boa Viagem}
\centerline{Niteroi 24210-340, RJ, Brazil}
\bigskip
\centerline{$^4$Institute for Theoretical Physics}
\centerline{Cologne University, D-50923 K\"oln, Euroland}
\bigskip
\bigskip
\bigskip
\parindent=0in
{\bf Abstract:}
The interacting lattice gas model is used to simulate fluid flow through an
open percolating porous medium with the fluid entering at the source-end 
and leaving from the opposite end. The shape of the steady-state 
concentration profile and therefore the gradient field depends on the  
porosity ($p$). The root mean square (rms) displacements of fluid
and its constituents (tracers) show a drift power-law behavior,
$R \propto t$ in the asymptotic regime ($t \to \infty$). The flux current
density ($j$) is found to scale with the porosity according to,
$j \propto {(\Delta p)}^{\beta}$ with $\Delta p = p - p_c$ and $\beta \simeq 1.7$. 
\bigskip


\newpage
\section{Introduction}
\bigskip
Motion of fluid/gas constituents (tracers) determines the collective 
diffusion of fluid/gas concentration. 
In general, tracers movements are correlated and depend 
on the concentration of fluid and the porosity of host medium/matrix
among other parameters such as temperature, interaction, pressure 
gradient, etc. 
Using a computer simulation model involving the interacting lattice gas,
we consider the flow of fluid through
a percolating porous medium [1] near its percolation threshold at a fixed
temperature. It involves the motion of constituents, global transport
of fluid, evolution of concentration profile in unsteady-flow, 
concentration gradient and flux rate in steady-state flow. 
A brief introduction and remark on some of these issues may help 
understanding our data. We simulate the flow through random walks
(stochastic motion of each particle), but in contrast to many previous
such diffusion studies we create a net flow by continuously injecting
new particles at the bottom, thus forcing the already existing ones to
move upwards since no two particles can occupy the same lattice site. The
system thus is spatially inhomogeneous in vertical direction.
\bigskip
\subsection{Concentration diffusion}
\bigskip
 
Diffusion of fluid concentration ($C$) in a homogeneous space is 
described by
$$\frac {\partial C} {\partial t} = D_m \nabla^2C \quad . \eqno(1)$$
Evolution of the concentration profile, diffusion of its front in 
unsteady flow, and drift in steady-state flow regimes are well 
understood [2]. 
How does the concentration diffuse (i.e., fluid flow) in a 
porous medium [3] such as percolating system [4-6]? What type of 
motion do the fluid and its constituents exhibit in unsteady and steady-state
flow regimes? How does the flux-rate depend on porosity? 
Addressing these questions becomes somewhat difficult with respect to 
solving the diffusion equation (1) particularly near percolation 
threshold [7-9] where the percolating pores are highly ramified; the boundary
conditions involved with the pore space become prohibitively
large to solve the diffusion equation numerically. We consider
an interacting lattice gas to model the fluid in order to address 
these questions in a percolating porous matrix.
The lattice gas consists of hard-core particles with a nearest
neighbor interaction (see section 2). 
\bigskip

\subsection{Tracer: diffusion, subdiffusion, drift}
\bigskip
A considerable progress has emerged in understanding the motion of a 
particle executing its random walk in percolating system [4,5].
Variation of the root mean square displacement ($R$) with time 
($t$) is one of the major quantities to characterize the type
of motion, i.e.,
$$R = A\cdot t^{\nu_1} + B \cdot t^{\nu_2} + ... \eqno(2)$$
If $\nu_1 > \nu_2$, then $R \simeq A\cdot t^{\nu_1}$ in the asymptotic
time limit ($t \to \infty$). With such a leading power-law 
variation, 
the motion is diffusive if $\nu_1 = 1/2$ and drift if $\nu_1 = 1$. 
In a percolating system at porosity (the fraction of pore sites)
above threshold, $p > p_c$, we have $\nu_1 = 1/2$. 
At the percolation threshold [7-9], on the other hand, 
the random walk motion becomes anomalous diffusion with 
$\nu_1 \simeq 0.2$ in three dimensions and $\nu_1 \simeq 0.3$ in 
two dimensions [10]. 
\bigskip

In presence of a biased field [11-13], the motion of a particle 
can still be described by above equation (2) except at very high bias 
with diffusion ($\nu_1 = 1/2$) in short time and drift ($\nu_2 = 1$) in 
the long time regime at $p \gg p_c$ with a crossover around $t = (A/B)^{2}$.
The motion becomes very complex [11] as the porosity is reduced
toward the threshold and the biased field competes
with the barriers at the pore boundaries.
The motion of a particle in such a porous medium has been 
extensively studied in a variety of biased fields for over a decade 
[11-17]. Many interesting findings have been reported
particularly due to increased computing power. Some of these results
include vanishing drift velocity above a characteristic or 
critical field [13,15], sub-diffusion and non-universal transport [11], 
log-periodic motion at high bias and large porosity [14].
Most of these studies deal with the motion of a single particle
in a biased field. Instead, we consider the flow of a fluid 
of interacting particles through
a percolating medium driven by concentration gradient (see below).
\bigskip

\subsection{Concentration gradient}
\bigskip
The mobile fluid constituents spread from high (toward the source end) 
to low particle concentration as the fluid flow from a source. The field
caused by the concentration gradient drives the fluid constituents as 
described by the concentration diffusion equation (1) in a homogeneous 
space. Stream of particles emanating from the source execute their stochastic 
(random walk) motion. 
The instantaneous distribution of particles forms a special morphology.
Particularly, the locus of the nearest neighbor particles on the moving
fluid front, i.e., the profile of the concentration front, leads to the morphology
of a percolating cluster at the percolation threshold. 
This is known as gradient percolation [18] since the dispersion and 
distribution of particles is caused by the concentration gradient. 
The shape of the concentration gradient is well known and the concentration 
of particles on the front provides a good estimate of the percolation 
threshold.
\bigskip

Evolution of the concentration profile of an interacting (nearest neighbor)
lattice gas in a homogeneous space [19] seems consistent with the diffusion 
equation (1). In fact, the velocity of the front from eq. (1) can be used 
to calibrate the time and length scale of the lattice gas simulations to 
understand the diffusion of specific system such as chlorine [19]. 
Obviously the motion of the front depends
on the shape of the concentration profile, i.e., the concentration gradient.
How does the concentration profile change if the
fluid moves through a percolating porous medium near its percolation threshold?
How does the front speed depend on the porosity near the threshold? 
\bigskip

Let us consider a finite system with the one end (bottom) connected to the
fluid source (particles) and the opposite end (top) open. The fluid moves 
from bottom to top as driven by the concentration gradient. The fluid
particles move from the source into the bottom and escape the system from the top. 
The concentration profile becomes stable as the system reaches the steady-
state, i.e., when the in-flux (at the bottom) becomes equal to the
out-flux 
(from the top). The steady-state concentration profile depends on porosity
and we investigate the changes in profile as we
vary the porosity near the threshold. Since the shape of the profile, i.e.
the concentration gradient, provides the driving field, the motion of the
front and the flux rate may depend on porosity as well. In this paper, 
we analyze some of these issues by a computer simulation model presented 
in next section (2) followed by results and discussion (in section 3). 
The conclusion is provided in the last section 4. 
\bigskip

\section{Model}
\bigskip
The host matrix is prepared on a simple-cubic $L \times L \times L$ lattice.
The percolating porous medium of porosity $p$ is generated by randomly 
distributing barriers, one at a site, on a fraction $p_b = 1-p$ sites.
The porosity is kept above the percolation threshold $p \ge p_c 
\; (\simeq 0.312)$ so that a spanning cluster of connected pore extends
from one of the sample to another. One end of this 
sample say the bottom ($x=0$) is connected to a source of fluid represented 
by mobile particles. The opposite end (top) is open so that the fluid 
particles entering the lattice at the bottom can escape from the top.
Each pore site in the bottom layer is occupied by the mobile fluid particle
(with one particle at a site, the excluded volume effect). 
\bigskip

The fluid is modeled by an interacting lattice gas. In order to introduce
interaction we assign an occupation variable ($S$) to each site. An empty
pore site $i$ is assigned $S_i = -1$ while a site $i$ with a fluid 
particle $S_i = 1$ and with a barrier $S_i = 0$. We use the following
interaction energy,

$$E = U \sum_{ij} S_i S_j \eqno(3)$$

where the interaction strength $U$ is set at unity in units of 
the Boltzmann constant ($k_B$). The summation is restricted to nearest
neighbor sites in this simulation. The fluid-fluid repulsive and fluid-pore
attractive interactions are thus considered. The fluid particles attempt to
move to an empty neighbor in a randomly selected direction with a Boltzmann distribution as in
the Metropolis algorithm at temperature $T$ [20]. A periodic boundary condition is used along the 
transverse ($y,z$) directions and open condition along the longitudinal
($x$) direction: a fluid particle cannot move below the bottom plane while
it can escape the system if it attempts to move above the top plane.
As soon as a particle leaves the pore site at the bottom, it is
occupied by another particle from the source. Thus a constant fluid 
concentration/density of unity is maintained in the bottom plane throughout
the simulation. An attempt to move each particles once is defined as one Monte 
Carlo step (MCS).
\bigskip

As the simulation proceeds, particles move out into the medium
from the bottom, fluid spreads, and concentration profile evolves.
The period during which none of the particles from the source reaches the 
top defines the unsteady ("short" time) regime. It is in this unsteady
state regime that the lattice gas simulation seems to reproduce the results
of continuum diffusion eq. (1) particularly the form of concentration profile 
and its motion in homogeneous space [19]. In the steady-state ("long" time) regime, 
the fluid-in-flux
(at bottom) equals the fluid-out-flux (at the top), and the continuity 
equation for the conservation of the mass is satisfied, i.e.,

$$\nabla \cdot j = 0 \quad . \eqno(4)$$
In the steady-state flow regime the concentration profile becomes stable.
One can evaluate the fluid current density ($j$) along the longitudinal 
direction.
We investigate the transport behavior of fluid (the center of
mass) and its constituent particles, the tracers and the flux response.
Simulations are carried out for a long time with many independent runs to obtain
a reliable estimate of physical quantities, particularly near percolation
threshold. Further, we have used different lattice sizes to check for  
significant finite size effects which are not detected in these data. 
\bigskip

\section{Results and Discussion}

Simulations are performed on different lattice sizes to look for 
finite size effects with most data generated on $30^3$ and $50^3$ samples.
The porosity is varied with $p \ge 0.312$. The temperature is constant
$T=2$ in units of Boltzmann constant. $N_r$ independent
samples, $N_r = 32 - 256$, are used via parallelization with MPI calls with
up to five million time steps. We have analyzed the variation of the rms 
displacement of tracers and their center of mass with time, concentration
profile, and flux rate density as a function of porosity near percolation
threshold. 
\bigskip
\subsection{RMS Displacements}
In section 1 we introduced the power-law dependence of the rms displacement
(eq. 2) for a single particle executing its stochastic motion. 
The fluid consists 
of many particles (in our model) and the collective motion of the fluid (i.e.,
the center of mass) results from the motion of individual particles, the 
tracers. The rms displacement of tracer, $R_t$, and that of the center of mass,
$R_c$, are described by the power-law dependence (eq. 2). Since the behavior of
$R_c$ describes the motion of the fluid, the center of mass of the particles
and fluid will be used synonymously as far as the fluid motion is concerned. 
As the fluid enters
the system from the bottom, the concentration gradient field drive the fluid 
from the bottom. The longitudinal ($x-$) component of the rms displacement
is much larger than the transverse ($y,z$) components. The power-law behavior
of the total rms displacement (for both tracer and center of mass of the fluid)
is dominated by the longitudinal component. Figure 1 shows the variation of
rms displacements for tracer with time at various porosities. 
The linear fit of data on a log-log scale in the short time regime
($10^2-10^4$ steps) suggest a power-law with exponent $\nu_1 \simeq 1/2$.
In the long time (asymptotic regime), the power-law exponent becomes higher
$\nu_2 \simeq 1$ leading to a drift-like motion. Such a crossover from
diffusion to drift is clearly seen at relatively higher porosities ($p_s = 
0.360, 0.400$). The crossover occurs at a characteristic time step ($t_c$) 
which increases on decreasing the porosity toward the threshold.
The rms displacement of the center of mass also shows similar crossover
behavior (figure 2). 
\bigskip

The closer to the percolation threshold we are, the longer ($t_c$) 
it takes to reach the asymptotic power-law behavior of the rms displacements
for the tracer and fluid. Particularly at $p = 0.312$, the lowest porosity 
we simulated near percolation threshold ($p_c \simeq 0.311608$) [7], 
it was difficult to reach long time power-law behavior with the 
larger samples. Therefore, we performed our simulation on a smaller sample 
($30^3$) for $5\times 10^6$ time steps.  The variation of the longitudinal 
component of the rms displacement with time is presented in figure 3 which 
shows a clear crossover to drift in the long time regime. Further we note that 
the qualitative behavior for the variation of the rms displacement with time 
remain the same as that on the larger sample for both tracer and fluid. 
\bigskip

It is worth pointing out that there are extensive computer simulations on
biased diffusion with a variety of biased fields [11-17]. In most of these 
studies
the tracer's motion slows down at low porosity (closer to percolation 
threshold) and high bias as the bias competes with the barriers at the pore
boundary. Thus there is no asymptotic drift behavior [15] of a biased random
walk in a percolating medium near threshold at high biased field.
In our driven system, the concentration gradient provides the bias. The
gradient field varies (see below) as we vary the porosity, but the asymptotic
drift behavior of the rms displacement of the tracer and fluid persists 
no matter how close we are to the percolation threshold. This implies that the
bias generated by the concentration gradient, perhaps not too strong, is an 
effective method of transport through a porous medium. 
\bigskip
\subsection{Concentration profile} 
\bigskip
Initially, the concentration profile is a delta function at $x=1$, since
there is no fluid particle in the sample except in the bottom plane
($x=1$) where each pore site is occupied by a fluid particle. 
As the simulation proceeds the fluid spreads and the profile changes.
Figure 4 shows a typical evolution of the concentration profile.
In the unsteady flow regime, i.e., when no fluid particle arrives at
the top, the shape of the profile seems consistent with the concentration
diffusion (eq. 1). In the long time (asymptotic) regime, the concentration
profile becomes stable and the system reaches a steady-state flow (see below).
The steady-state profile depends on porosity. Figure 5 shows the steady-state
profile at various porosities above the percolation threshold. At low porosity
($p = 0.312$), we see a spatial variation in concentration gradient.
A relatively linear gradient ($dc/dx$) develops at higher porosities.
The gradient field (concentration gradient) is delicately controlled by
porosity in such a way that the asymptotic drift motion of tracer and fluid
occurs. The net transfer of fluid across the sample, i.e., the flow response,
depends on porosity which is discussed in the following.
\bigskip
\subsection{Flow response}
\bigskip
Fluid enters the sample at the bottom ($x=1$) and leaves from the top ($x=L$).
There is a net fluid transfer ($q$) along $x-$direction. We evaluate the
current density ($j$) resulting from the rate of fluid (mass) transfer,
$$j = \frac {1}{L^2} \cdot \frac {dq}{dt} \eqno(5)$$
In fact, we evaluate both, the input (bottom) and output (top) current 
densities ($j$) separately. Figure 6 shows the variation of these current
(flux) densities at various porosities. As expected, the input flux density
is higher and output flux density is lower initially, i.e. in the 
unsteady-flow regime. In steady-state, both must be equal, as seen in 
figure 6. Thus we see that the flow has reached the steady-state at nearly
all porosities except very close to the percolation threshold ($p_s=0.312$).
Looking at the trend in data, it is easy to estimate the flux density by
extrapolation. However, we have carried out our simulation for a long time
on a smaller sample to achieve the steady state at $p_s=0.312$; figure 7
shows the input and output flux density.
\bigskip

It would be interesting to quantify how the steady-state flux density ($j$)
depend on porosity. Figure 8 shows the variation of $j$ with porosity ($p_s$).
We see a relatively good power-law dependence,
$$j \propto {(\Delta p)}^{\beta} \eqno(6)$$
where $\Delta p = p_s - p_{c}$ and $\beta \simeq 1.7$. 
Fig. 8 shows that this effective exponent $\beta$ increases with increasing $L$ and thus asymptotically may be compatible with the exponent $\simeq 2$
of random resistor networks [3-6].

\bigskip
\section{Conclusion}
\bigskip
A computer simulation model is used to study the density profile and fluid
flow through an open porous medium near the percolation threshold. The fluid is
driven by the concentration gradient which is evolved from maintaining a
constant concentration, unity at the source (bottom) and zero at the top.
RMS displacements of tracer and fluid are studied in detail at various
porosities near the percolation threshold. The long time asymptotic power-law
behavior of the rms displacement $R \propto t$ is found at all porosities above
the threshold unlike the biased diffusion in previous simulations [11-15].
We believe that such an asymptotic drift is achieved due to the 
concentration gradient caused by a delicate competition between the stochastic
motion of the fluid particles and the pore barriers. The shape of the
density profile is sensitive to porosity.
\bigskip

We have shown that the fluid flux through such a percolating porous medium 
reaches a steady-state above the percolation threshold.
The flux density ($j$) depends on porosity and we characterize it by
an empirical power-law relation (eq. 6). We plan to study the effects of 
parameters such as temperature, pressure, etc. in such a gradient driven
system. 

\bigskip

{\bf Acknowledgment:}
This work is supported under ONR PE\#
0602435N. This work was supported in part by a grant of computer 
time from the DOD High Performance Computing Modernization 
Program at the Major Shared Resource Center (MSRC), NAVO, 
Stennis Space Center.
\newpage

\leftline{\bf References:}
\bigskip

[1] H.E. Stanley and J.S. Andrade Jr., Physica A 295, 17 (2001).
\smallskip

[2] E.L. Cussler, {\em ``Diffusion: Mass transfer in fluid systems''}
(Cambridge University Press, 1984); 
C.A. Silebi and W.E. Schiesser, {\em ``Dynamic Modeling of Transport
Process Systems''} (Academic Press, Inc., 1992). 
\smallskip

[3] M. Sahimi, {\em ``Flow and transport in Porous Media and Fractured
Rock''} (VCH Weinheim, 1995). 
\smallskip

[4] D. Stauffer and A. Aharony, {\em ``Introduction to Percolation Theory''},
Second Edition (Taylor and Francis, 1994).
\smallskip

[5] A. Bunde and S. Havlin, eds., {\em ``Fractals and Disordered Systems''},
Second Edition (Springer, New York, 1996).
\smallskip

[6] M. Sahimi, {\em ``Application of Percolation Theory''} (Taylor and 
Francis, 1994). 
\smallskip

[7] G. Paul, R.M. Ziff, and H.E. Stanley, LANL eprint, cond-mat/0101136 (2001).
\smallskip

[8] C.D. Lorenz and R.M. Ziff, Phys. Rev. E 57, 230 (1998).
\smallskip

[9] D. Stauffer and R.M. Ziff, Int. J. Mod. Phys. C 11, 205 (2000).
\smallskip

[10] S. Havlin and D. Ben Avraham, Adv. Phys. 36, 395 (1987).
\smallskip

[11] R.B. Pandey, Phys. Rev. B 30, 489 (1984).
\smallskip

[12] H. Boettger and V.V. Bryskin, Phys. Stat. Sol. (b) 113, 9 (1982).
\smallskip

[13] M. Barma and D. Dhar, J. Phys. C 16, 1451 (1983).
\smallskip

[14] D. Stauffer and D. Sornette, Physica A 252, 271 (1998).
\smallskip

[15] D. Dhar and D. Stauffer, Int. J. Mod. Phys. C 9, 349 (1998).
\smallskip

[16] E. Seifert and M. Sussenbach, J. Phys. A 17, L 703 (1994).
\smallskip

[17] A. Bunde, S. Havlin, and H.E. Roman, Phys. Rev. A 42, 6274 (1990).
\smallskip

[18] M. Rosso, B. Sapoval, and J.-F. Gouyet, Phys. Rev. Lett. 57, 3195 (1986);
R.P. Wool, {\em ``Polymer Interfaces''}, Chapter 4 (Hanser Publishers, 1995).
\smallskip

[19] R.B. Pandey, W. Wood, and J.G. Gettrust, Preprint (2001).
\smallskip

[20] K. Binder, ed., {\em `` The Monte Carlo Methods in Condensed Matter
Physics''} Topics in Applied Physics, Vol. 71 (Springer-Verlag, 1995).
\bigskip

\newpage


\begin{figure}[hbt]
\begin{center}
\includegraphics[angle=-00,scale=0.6]{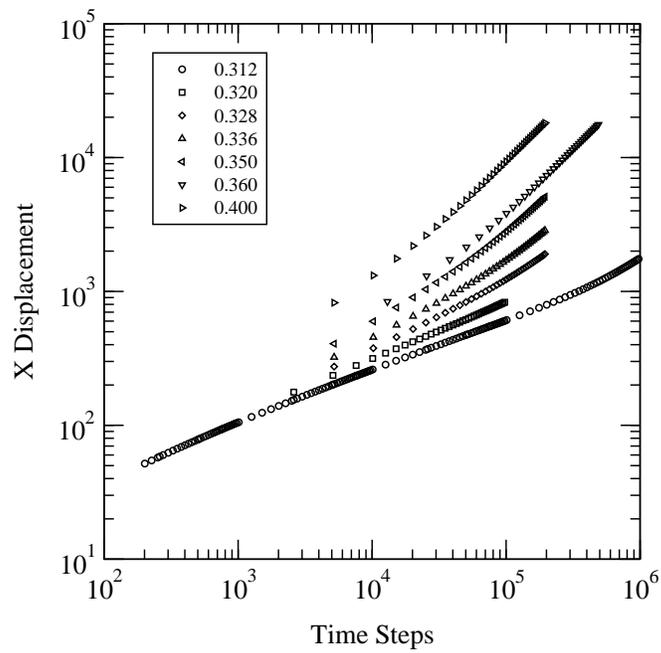}
\end{center}
\caption{
Root mean square (rms) displacement of the tracer with time steps on a log-log
scale for various porosities, $p = 0.312 - 0.4$ on a $50^3$ sample with
$T = 2.0$. $N_r = 32-256$ independent samples.
}
\end{figure}
\bigskip

\begin{figure}[hbt]
\begin{center}
\includegraphics[angle=-00,scale=0.6]{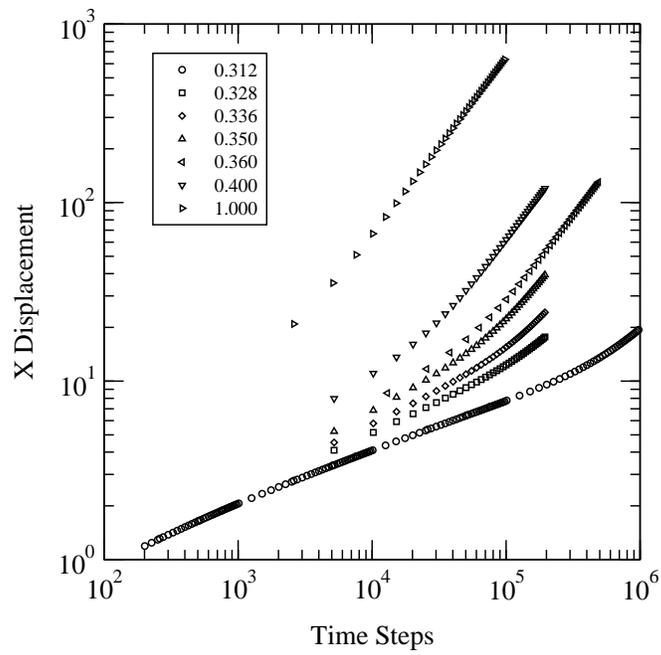}
\end{center}
\caption{
RMS displacement of the center of mass of the fluid particles with time
steps. Statistics is the same as in figure 1.
}
\end{figure}
\bigskip

\begin{figure}[hbt]
\begin{center}
\includegraphics[angle=-00,scale=0.6]{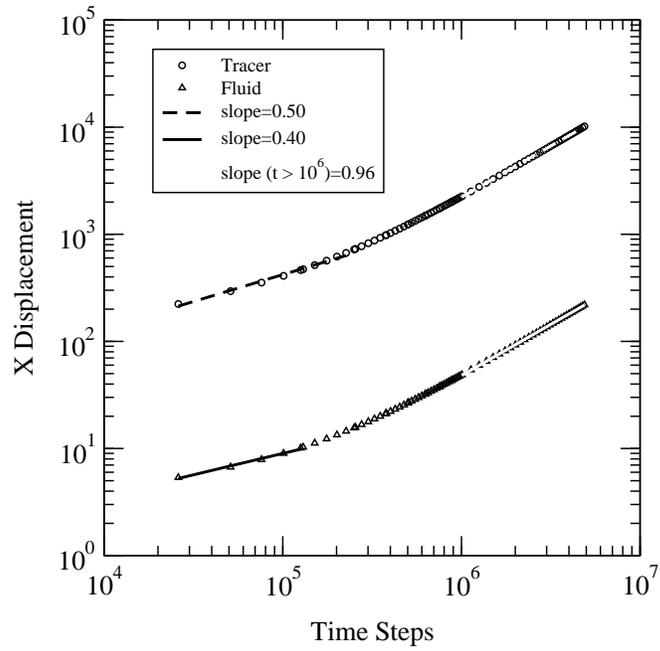}
\end{center}
\caption{
RMS displacement of the tracer and the center of mass versus time steps
at $p = 0.312$ on a $30^3$ sample with $N_r=256$. 
}
\end{figure}
\bigskip

\begin{figure}[hbt]
\begin{center}
\includegraphics[angle=-00,scale=0.6 ]{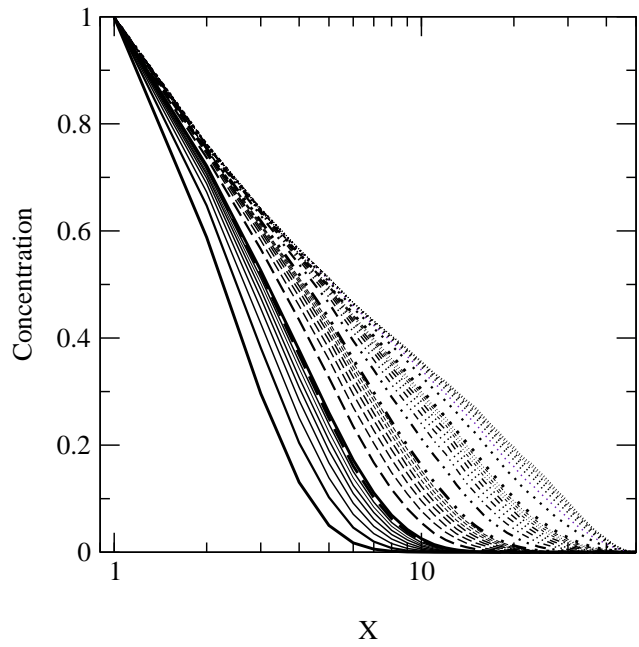}
\end{center}
\caption{
Evolution of concentration profile on a $50^3$ sample at $p=0.312$ with
$N_r=256$.
}
\end{figure}
\bigskip

\begin{figure}[hbt]
\begin{center}
\includegraphics[angle=-00,scale=0.6]{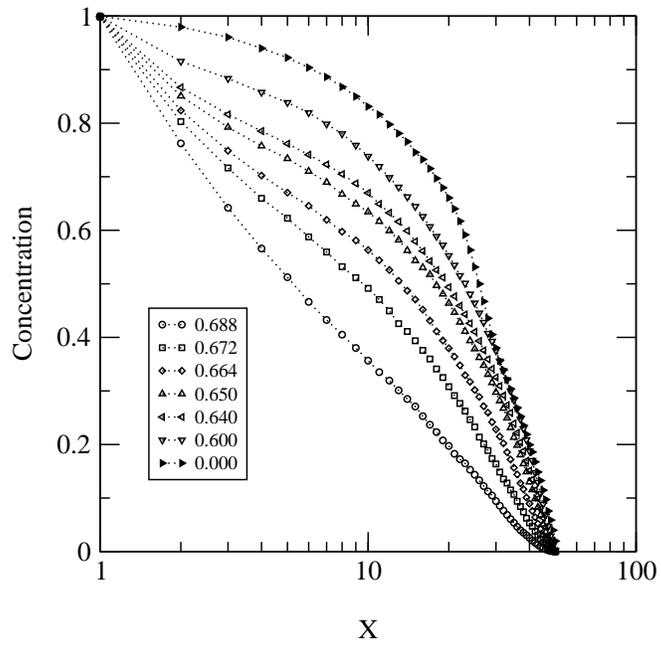}
\end{center}
\caption{
Concentration profile in steady-state. Statistics is the same as in figure 1.
}
\end{figure}
\bigskip

\begin{figure}[hbt]
\begin{center}
\includegraphics[angle=-00,scale=0.6]{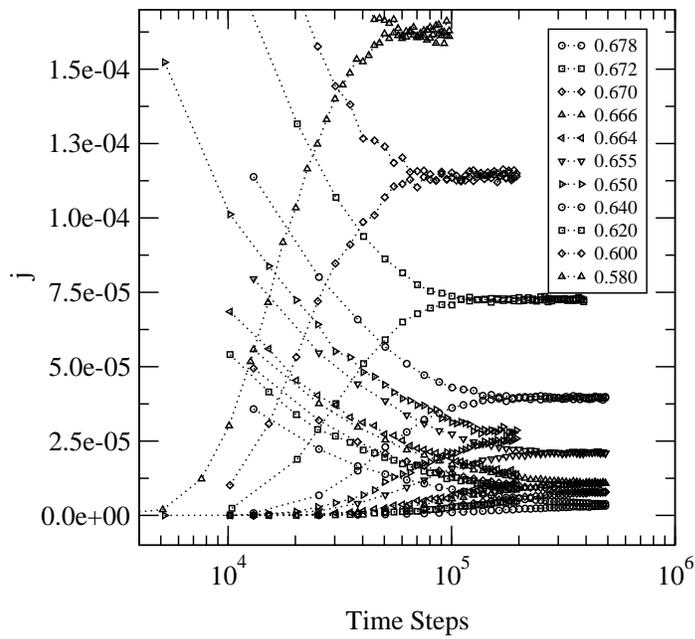}
\end{center}
\caption{
Flux rate density ($j$) versus time steps at various porosities. Statistics
is the same as in figure 1. The top data is the flux-in at the bottom and
bottom data is the flux-out from the top.
}
\end{figure}
\bigskip

\begin{figure}[hbt]
\begin{center}
\includegraphics[angle=-00,scale=0.6]{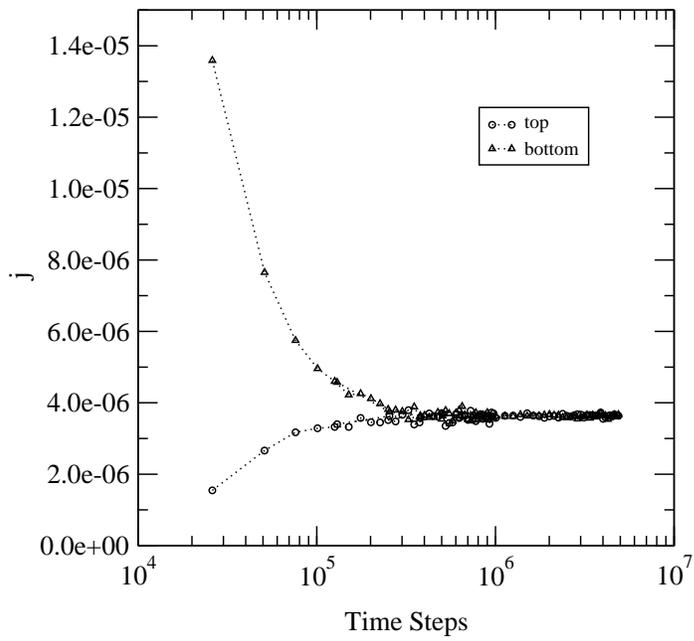}
\end{center}
\caption{
Flux rate density ($j$) versus time steps at $p = 0.312$ on a $30^3$ sample
with the same statistics as in figure 3.
}
\end{figure}
\bigskip

\begin{figure}[hbt]
\begin{center}
\includegraphics[angle=-90,scale=0.6]{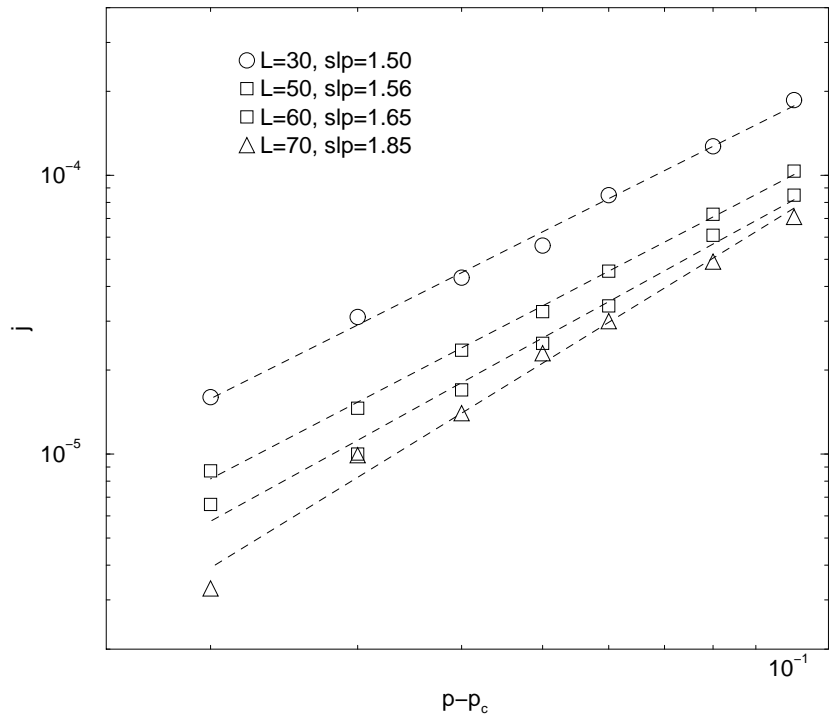}
\end{center}
\caption{
Log-log plot of $j$ versus $p - p_c$ at $T=1$ for $L = 30$ ($\times$), 50 (+), 
60 (stars), and 70 (squares), with slopes 1.5, 1.6, 1.7, 1.75, respectively. We
average over usually 8 lattices and $10^5$ time steps. This flux rate density
$j$ varies as the concentration gradient $\propto 1/L$. 
}
\end{figure}
\end{document}